\documentclass[preprint,aps,12pt,showpacs,nofootinbib,tightenlines]{revtex4}
\usepackage{amsmath}
\usepackage{amssymb}
\usepackage{epsfig}
\usepackage{graphicx}
\usepackage{ctex}

\begin{document}
\def\pslash{\rlap{\hspace{0.02cm}/}{p}}
\def\eslash{\rlap{\hspace{0.02cm}/}{e}}
\title{ Lepton flavor violating Higgs couplings and single production of the Higgs boson via $e \gamma$ collision}
\author{Chong-Xing Yue , Cong Pang and Yu-Chen Guo\\
{\small Department of Physics, Liaoning  Normal University, Dalian
116029, P. R. China}
\date{\today}
\vspace*{1.5cm}}

\begin{abstract}
Taking into account of the constraints on the lepton flavor violation (LFV) couplings of the standard model (SM) Higgs boson H with leptons from low energy experiments and the recent CMS results, we investigate production of the SM Higgs boson associated with a lepton $\tau$ via $e\gamma$ collision at the ILC and LHeC experiments. The production cross sections are calculated, the LFV signals and the relevant SM backgrounds are examined. The LFV signals of the SM Higgs boson might be observed via $e\gamma$ collision in future ILC experiments.

\end{abstract}

\email{cxyue@lnnu.edu.cn}

\maketitle
\newpage

\noindent{\bf 1. Introduction }\vspace{0.5cm}

During the past decades, neutrino oscillation experiments have provided us with very convincing evidence that neutrinos are massive particles mixing with each other [1, 2], which means that lepton flavor is not an exact symmetry and lepton flavor violating (LFV) processes exist in nature. However, in the standard model (SM), these processes are strongly suppressed by GIM mechanism, making them unobservable at current or planned experiments. Thus, LFV processes provide one of the most interesting probes to physics beyond the SM and the detection of any LFV process would provide a clear evidence of new physics. This fact gives us a strong motivation to search for charged LFV. For example, the LHC has given some of upper limits on the lepton number violation [3] though at this moment more stringent limits are given by Belle Collaboration.

At present, it is widely believed that the recently discovered scalar particle at the LHC [4, 5] behaves as the SM Higgs boson related to the mechanism of electroweak symmetry breaking. The newly values of its mass are M$^{ATLAS}_{H}$ = 125.5$\pm$0.6 GeV [6] and M$^{CMS}_{H}$ = 125.7$\pm$0.4 GeV [7] measured by the ATLAS and CMS collaborations, respectively. With the Higgs boson discovery, particle physics enters a new era of detailed and careful study of its properties, such as couplings and decays. Accurate understanding of the Higgs properties together with new data from high energy collider experiments will provide us a tool for exploring new physics.

A complementary and well motivated means for investigating the Higgs boson properties is the search for its non-SM properties. Among these, LFV Higgs couplings form an interesting class [8, 9]. Although strictly forbidden at tree level in the SM, the Higgs LFV couplings arise naturally in many well-motivated extensions of the SM. Discovery of the Higgs boson at the LHC has caused renewed interest in considering LFV effects associated with this scalar particle [8, 9, 10, 11]. Searching for the LFV Higgs couplings at the LHC offer an interesting possibility to test for new physics effects that might has escaped current experimental constraints.

 Furthermore, observing the LFV signals has become experimentally available. The CMS collaboration has recently reported a slight excess with a significance of 2.4$\sigma$ in the search for
the LFV decay $H\rightarrow \mu\tau$ [12]:
 \begin{eqnarray}
Br(H\rightarrow \mu\tau)=(0.84_{-0.37}^{+0.39})\%,
\end{eqnarray}
 where the final state is a sum of $\mu^{+}\tau^{-}$ and $\mu^{-}\tau^{+}$. Certainly, this recent hint, which although has received  amount of attention in the literature [13, 14], needs to be confirmed  or rejected with more data by both ATLAS and CMS experiments at the LHC run II.

The proposed international linear collider (ILC) [15], which is an $e^{+}e^{-}$ collider with high energy and luminosity, has particularly clean environment and will provide an opportunity for high precision measuring various observables related the SM Higgs boson, gauge bosons and fermions, and further detecting new physics effects. Such a machine is well suited to an in-depth analysis of elementary particle interactions within and beyond the SM. The potential of the ILC can be further enhanced by considering $\gamma \gamma$ and $e \gamma$ collisions with the photon beam generated by the backward Compton scattering of incident electron- and laser-beams [16, 17]. The high energy $\gamma \gamma$ or $e \gamma$ collider might provide us a good chance to precision test the SM and further to search new particles.

The proposed large hadron electron collider (LHeC) can be realized by colliding the existing 7 TeV proton beam with E$_{e}$ = 50 $\sim$ 200 GeV electron (positron) beam and its anticipated integrated luminosity is about at the order of 10 $\sim$ 100 fb$^{-1}$ [18]. The LHeC can provide better condition for studying a lot of phenomena comparing to the ILC due to the high center-of-mass (c. m.) energy and to the LHC due to more clear environment. Thus, it may play a significant role in the discovery of new physics beyond the SM. The LHeC can also be transformed to $e\gamma$ collision with the photon beam radiated from proton and the radiating proton remaining intact; thus providing an extra experimental handle(forward proton tagging) to help reduce the background [19]. Despite a lower available luminosity, $e\gamma$ collision occurs under better known initial conditions, with fewer final states and thus can be studied as a complementary tool to normal $ep$ collision at the LHeC. In this paper, we investigate single production of the SM Higgs boson H via $e \gamma$ collision processes $e \gamma$ $\rightarrow$ $\ell H$ ($\ell$ = $\mu$ or $\tau$) at the ILC and LHeC, which are induced by the LFV couplings $He\ell$, and discuss the possibility of detecting its LFV effects.

The layout of the present paper is as follows. Taking into account of the constraints from the LFV processes $\ell_{i}\rightarrow\ell_{j} \gamma$ and $\ell_{i}\rightarrow\ell_{j} \ell_{k} \ell_{l}$ on the LFV Higgs couplings $H\ell_{i} \ell_{j}$, single production of the SM Higgs boson H via $e \gamma$ collision at the ILC and LHeC are calculated in sections 2 and 3, respectively. The relevant signals and backgrounds are discussed in these two sections. Our conclusions are given in section 4.

\vspace{0.5cm} \noindent{\bf 2. LFV production of the SM Higgs boson H via $e \gamma$ collision at the ILC }

\vspace{0.5cm}In the mass basis, the couplings of the Higgs boson to charged leptons can be general written as
\begin{eqnarray}
\mathcal{L}=-Y_{ij}H\overline{\ell^{i}_{L}}\ell^{j}_{R}+h.c.,
\end{eqnarray}
where $i,j =e,\mu,\tau$ and in the SM $Y_{ij } = m_{\tau}/\upsilon \delta_{ij}$ with $\upsilon$ =246 GeV. The precision measurement data and the experimental upper limits on some LFV processes can give constraints on the Yukawa couplings $Y_{ij}$. The strongest low-energy constraints on the couplings $Y_{\mu e},Y_{\tau \mu}$ and $Y_{\tau e}$ come from the experimental upper limits on the LFV processes $\mu \rightarrow e \gamma$, $\tau \rightarrow \mu \gamma$ and $\tau \rightarrow e \gamma$ [20]. References [8, 9] have  shown that the constraint on $g_{\mu e}$ is much stronger, and require the branching ratio $Br(H \rightarrow \mu e)$ to be smaller than $2 \times 10^{-8}$, which is not likely to be observed at the LHC. While the constraints on the LFV Higgs couplings $g_{\tau \mu}$ and $g_{\tau e}$ are weaker, allowing for the branching ratio $Br(H \rightarrow \tau \mu)$ or $Br(H \rightarrow \tau e)$ as high as $\sim 10\%$, which is comparable to $Br(H \rightarrow \tau \tau)$ in the SM. We will therefore not consider single production of the SM Higgs via the LFV process $e \gamma \rightarrow \mu H$ in this paper. So, only the LFV coupling $H \tau e$ is related our calculation. In our numerical estimation, we will assume $\sqrt{|Y_{\tau e}|^{2}+|Y_{e \tau}|^{2}}$ $\leq$ 0.014 [9].

Single production of scalar particles with the LFV couplings $S\ell_{i} \ell_{j}$ via $e \gamma$ collision at the ILC has been discussed in SUSY and topcolor theories [21]. From above discussions, we can see that the SM Higgs boson H can be produced in association with a lepton $\tau $ via $e \gamma$ collision mediated by the LFV coupling $H \tau e$, as shown in Fig.1. The differential cross section for the subprocess $e \gamma \rightarrow \tau H$ is expressed by
\begin{eqnarray}
{\frac{\mathrm{d}\hat{\sigma}(\hat{s},P_{e })}{\mathrm{d}\cos\theta}}=\frac{\alpha_{e}\sqrt{\lambda(\frac{m_{\tau}^{2}}{\hat{s}},\frac{m_{H}^{2}}{\hat{s}})}[Y_{\tau e }^{2} + Y_{e \tau}^{2}][A_{-}(A_{+}^{2} + 4 B^{2}) - \frac{16 B m_{\tau}^{2}}{\hat{s}}]}{32 \hat{s} A_{-}^{2}},
\end{eqnarray}
where $\alpha_{e}$ is the fine-structure constant, $\hat{s}$ is c. m. energy of the subprocess $e \gamma \rightarrow \tau H $ and $\theta$ is the scattering angle of the outgoing lepton $\tau $ from the beam direction. $A_{\pm}=1 + B \pm \lambda^{1/2}(\frac{m_{\tau}^{2}}{\hat{s}},\frac{m_{H}^{2}}{\hat{s}})\cos\theta$ with $B=\frac{m_{\tau}^{2}-m_{H}^{2}}{\hat{s}}$ and $\lambda(a,b) = 1+a^{2}+b^{2}-2a-2b-2ab$.

\begin{figure}[htb]
\vspace{-0.2cm}
\begin{center}
 \epsfig{file=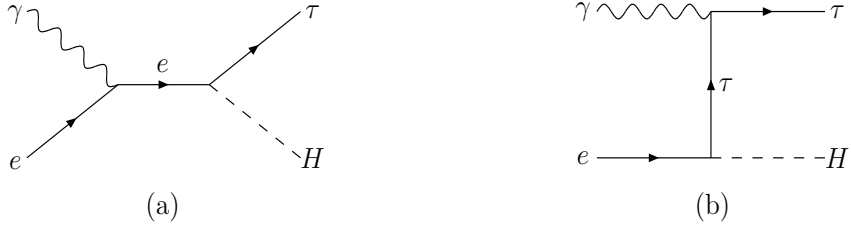, width=320pt,height=85pt}
 \vspace{-0.5cm}
 \caption{ The Feynman diagrams for the subprocess $e \gamma \rightarrow \tau H$.}
 \label{ee}
\end{center}
\end{figure}

After calculating the cross section $\hat{\sigma}(\hat{s})$ for the subprocess $e \gamma \rightarrow \tau H$, the effective cross section $\sigma (s)$ at the ILC can be evaluated from $\hat{\sigma}(\hat{s})$ by convoluting with the photon structure function $f_{\gamma/e}(x)$ as

\begin{eqnarray}
\sigma(s)=\int_{(m_{H}+m_{\tau})^{2}/s}^{x_{max}}\mathrm{d}xf_{\gamma/e}(x)\int_{(\cos\theta)_{min}}^{(\cos\theta)_{max}}\mathrm{d\cos\theta}\frac{\mathrm{d}\hat{\sigma}(\hat{s},P_{e })}{\mathrm{d}\cos\theta}.
\end{eqnarray}
Where $x_{max} = \xi / (1 + \xi)$, $x=\hat{s}/s$, in which $\sqrt{s}$ is the c. m. energy of the ILC experiments. In order to avoid producing $e^{+}e^{-}$ pairs by the interaction of the incident and backscattered photons, there should be $\xi \leq 4.8$. For $\xi = 4.8$, there is $x_{max} \approx 0.83$. The photon distribution function $f_{\gamma/e}(x)$ can be expressed as [16]
\begin{eqnarray}
f_{\gamma/e}(x)=\frac{1}{D(\xi)}\{(1 - x) + \frac{1}{1 -x} - \frac{4 x}{\xi (1 - x)} + \frac{4 x^{2}}{\xi^{2} (1 - x)^{2}}\}
\end{eqnarray}
with
\begin{eqnarray}
D(\xi)=(1 - \frac{4}{\xi} - \frac{8}{\xi^{2}}) \ln(1 + \xi) +\frac{1}{2} + \frac{8}{\xi} - \frac{1}{2 (1 + \xi)^{2}}.
\end{eqnarray}
It is known that the forward and backward directions in $e \gamma$ collision are blind spots for detection of scattered particles. To make the scattered particles be detected, we impose the cut $10^{\circ} \leq \theta \leq 170^{\circ}$.

\begin{figure}[htb]
\vspace{-0.5cm}
\begin{center}
 \epsfig{file=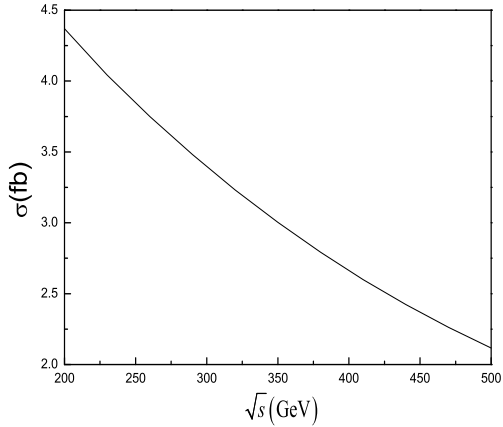, width=220pt,height=200pt}
 \vspace{-0.5cm}
  \caption{The production cross section $\sigma$ of the subprocess $e \gamma \rightarrow \tau H$ at the ILC as a
   \hspace*{1.6cm} function  of the center-of-mass (c. m.) energy $\sqrt{s}$ for $\sqrt{|Y_{\tau e}|^{2}+|Y_{e \tau}|^{2}}$ = 0.014.}
 \label{ee}
\end{center}
\end{figure}

It is obvious that the effective production cross section $\sigma$ of the subprocess $e \gamma \rightarrow \tau H$ depends the two free parameters: the c. m. energy $\sqrt{s}$ and the LFV Higgs coupling $H \tau e $. Although some new physics models can not explain the maximal value of the $H \tau e $ coupling given by the experimental upper bounds for the LFV process $\tau \rightarrow e \gamma$, it is theoretical possible. As numerical estimation, we fix  $\sqrt{|Y_{\tau e}|^{2}+|Y_{e \tau}|^{2}}$ = 0.014. Our numerical results are showed in Fig.2, in which we plot the effective cross section  $\sigma$ as a function of the c. m. energy $\sqrt{s}$ for $M_{H} = 125 $ GeV and $m_{\tau} = 1.777 $ GeV. One can see from Fig.2 that, in the case of considering the bound from the process $\tau \rightarrow e \gamma$ on the LFV Higgs coupling $H \tau e $, the values of the cross section $\sigma$ can reach $4.4 \sim 2.1$ fb for $\sqrt{s}$ = 200 $\sim$ 500 GeV. For the SM Higgs boson with $M_{H} = 125 $ GeV, the decay process $H \rightarrow b \overline{b} $ is a main decay channel and its branching ratio is about 58\% [22]. Then, the signal final state can be seen as $\tau b \overline{b}$ for the LFV Higgs production via $e \gamma$ collision at the ILC. For $\sqrt{s}$ = 200 $\sim$ 500 GeV, there will be 1496 $\sim$ 714 $\tau b \overline{b}$ events to be generated in the future ILC experiment with the integrated luminosity $\mathcal{L}$$_{int}$ = 340 fb$^{-1}$.

The main background for the signal state $\tau b \overline{b}$ comes from the process $e \gamma \rightarrow W Z \nu \rightarrow \tau b \overline{b} \nu \overline{\nu}$. Certainly, other processes, such as the SM process  $e \gamma \rightarrow W\gamma \nu \rightarrow \tau b \overline{b} \nu \overline{\nu}$, can also contribute to the background, while their contributions are much smaller than those of  $e \gamma \rightarrow W Z \nu$. To more exactly calculate the background, we use MadGraph5 [23] to write down all the tree Feynman diagrams for $e \gamma \rightarrow \tau b \overline{b}(\nu \overline{\nu})$ and to calculate the contributions to the background cross section. In our numerical estimation, similarly as above, we use the spectrum of photons obtained by the laser backscattering technique [16, 17], which is embedded in MadGraph. We find that, for $\sqrt{s}$ = 200 $\sim$ 500 GeV, the background cross section is in the range of 1.6$\times 10^{-4}$ $\sim$ 1.9$\times 10^{-1}$ fb, before any kinematic cuts applied, which is small enough compare to the signal cross section. The backgrounds can strongly be suppressed by the invariant mass cut for $b\overline{b}$. If we further assume that the tau lepton decays into various hadronic and leptonic modes, the signal and the relevant backgrounds would become complex. However, the conclusion that the signal cross section is much larger than that for the relevant backgrounds is not changed.

\begin{figure}[htb]
\vspace{-0.5cm}
\begin{center}
 \epsfig{file=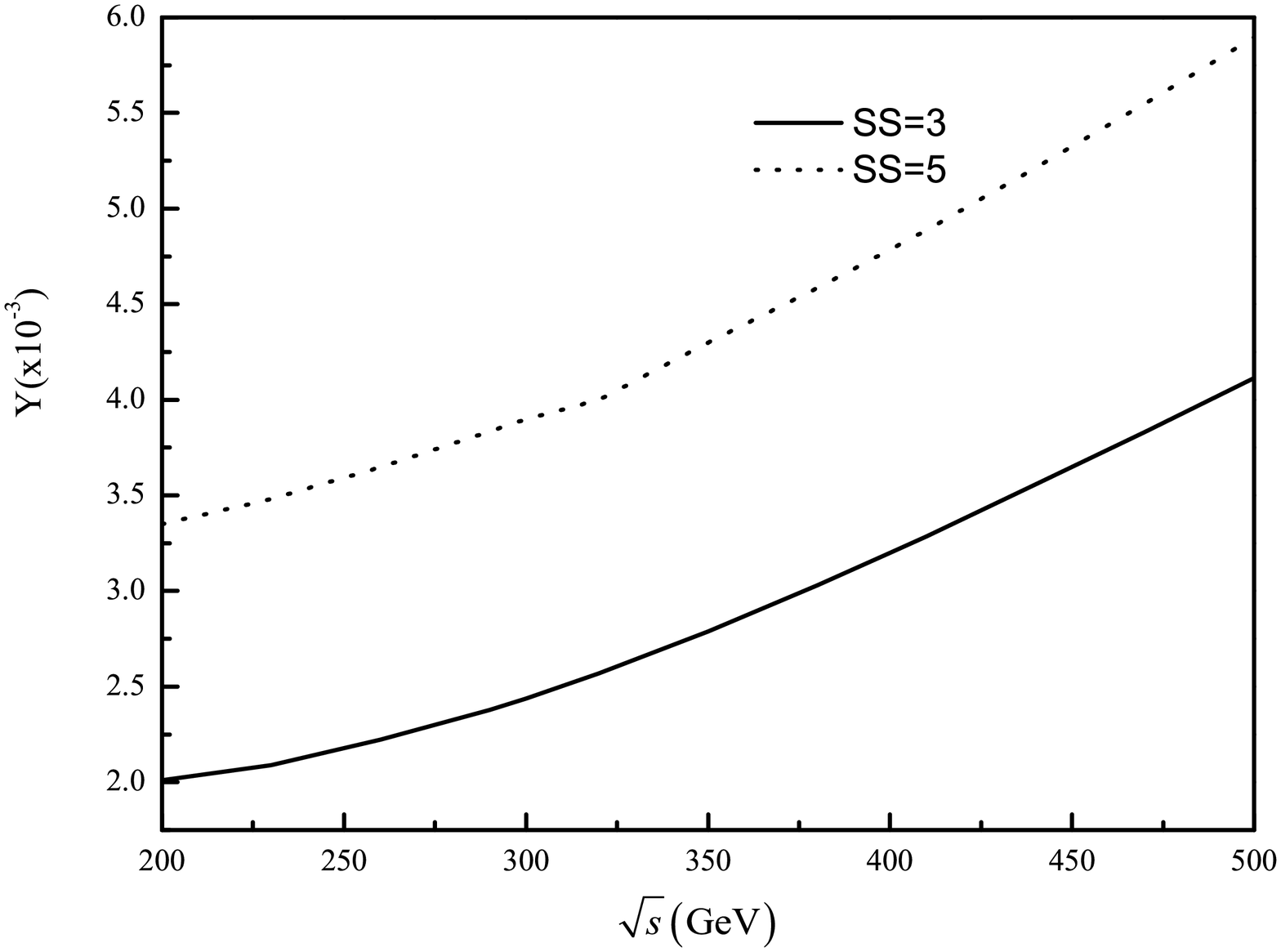, width=220pt,height=200pt}
 \epsfig{file=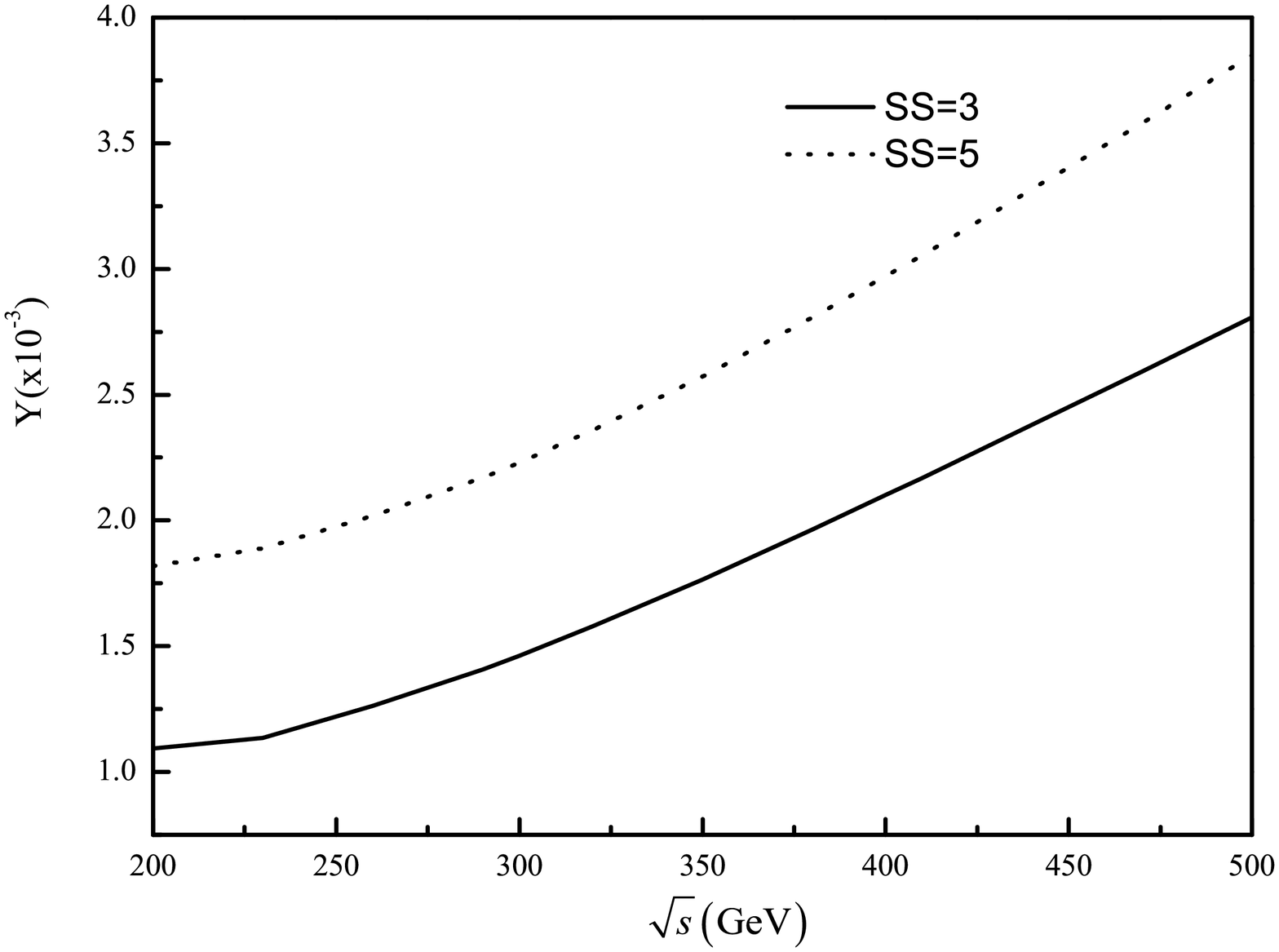, width=220pt,height=200pt}
 \vspace{-0.5cm}
  \caption{ Variation of the LFV $H \tau e$ coupling  $Y$ with  the c. m. energy  $\sqrt{s}$ for the ILC experiments \hspace*{1.55cm}with the integrated luminosity $\mathcal{L}$$_{int}$ = 100 fb$^{-1}$(left) and 340 fb$^{-1}$(right).}
 \label{ee}
\end{center}
\end{figure}

For the  statistical significance, we use the definition $SS=S/\sqrt{S+B}$, where S and B denote the number of signal and background events. It is obvious that $SS$ is a function of two parameters, namely  the c. m. energy $\sqrt{s}$ and the LFV $H \tau e$ coupling  $Y=\sqrt{|Y_{\tau e}|^{2}+|Y_{e \tau}|^{2}}$. Performing the scanning over the parameter space we can derive the experimental evidence region ($SS\geq3$) and experimental discovery region (($SS\geq5$). The results are shown in Fig.3. One can see from this figure that, as long as the value of the coupling parameter  $Y=\sqrt{|Y_{\tau e}|^{2}+|Y_{e \tau}|^{2}}$ is larger than $1\times 10^{-3}$, the future ILC experiment with the integrated luminosity $\mathcal{L}$$_{int}$ = 340 fb$^{-1}$ will produce the experimental evidence for the LFV Higgs coupling $H \tau e $.

Using the CMS results in the search for the LFV decay $H\rightarrow \mu\tau$ and other existing stringent experimental limits, Ref.[14] has studied the constraint on the coupling LFV $H \tau e$. The CMS result $\sqrt{|Y_{\tau \mu}|^{2}+|Y_{\mu \tau}|^{2}}<3.6\times 10^{-3}$ [12] can give the constraint $Y=\sqrt{|Y_{\tau e}|^{2}+|Y_{e \tau}|^{2}}<2.0\times 10^{-3}$.  So, even if the CMS result is confirmed in near future, the signals of the LFV subprocess $e \gamma \rightarrow \tau H$ might be detected in future ILC experiments.

\vspace{0.5cm} \noindent{\bf 3. LFV production of the SM Higgs boson via $e \gamma$ collision at the LHeC}
\vspace{0.5cm}

The LHeC is a future electron-proton collider and being planned at CERN [18]. Improvement in the precision determinations of parton distribution functions (PDFs) and the strong coupling constant $\alpha_{s}$, the LHeC would allow us to predict new particle production cross sections with sufficient accuracy to distinguish between different explanations of new physics phenomena. At the LHeC, a quasi-real photon (with low virtuality $ Q^{2} = -q^{2}$) can be emitted from the proton, which is named intact or forward proton and will be detected by the forward detectors with very large pseudorapidity. The forward proton detectors are planned to be built at about $220 m$ from the ATLAS main detector within the AFP project [24] and the CMS
and TOTEM collaborations plan to use their forward proton detectors located at about the same position (CT-PPS project)[25]. Using these detectors, the relevant photon interactions
can be detected with an unprecedented precision.

The emitted quasi-real photons can be described in the framework of equivalent photon approximation (EPA) and show a spectrum of virtuality $Q^{2}$ and the energy $E_{\gamma}$ [26]

\begin{eqnarray}
\frac{\mathrm d N_{\gamma}}{\mathrm dE_{\gamma}\mathrm dQ^{2}}=\frac{\alpha_{e }}{\pi}\frac{1}{E_{\gamma}Q^{2}}[(1-\frac{E_{\gamma}}{E})(1-\frac{Q_{min}^{2}}{Q^{2}})F_{E} + \frac{E_{\gamma}^{2}}{2E^{2}} F_{M}]
\end{eqnarray}
with
\begin{eqnarray}
Q_{min}^{2} =\frac{M_{p}^{2}E_{\gamma}^{2}}{E(E-E_{\gamma})},   F_{E}=\frac{4 M_{p}^{2}G_{E}^{2} + Q^{2}G_{M}^{2}}{4M_{p}^{2} + Q^{2}},
\end{eqnarray}
\begin{eqnarray}
G_{E}^{2}=\frac{G_{M}^{2}}{\mu_{p}^{2}}=(1 + \frac{Q^{2}}{Q_{0}^{2}})^{-4}, F_{M} = G_{M}^{2}, Q_{0}^{2} = 0.71 GeV^{2}.
\end{eqnarray}
Where $M_{p}$ is the mass of the proton, $E$ is the energy of the incoming proton beam, which is related to the photon energy by $E_{\gamma}=\xi E$. The parameter $\xi$ indicates the fractional proton momentum loss and is also  defined as the forward detector acceptance $\xi = \Delta E/ E$, in which $\Delta E$ is the loss energy of the emitted proton beam. $\mu_{p}^{2}$ = 7.78 is the magnetic moment of the proton, $F_{E}$ and $F_{M}$ are functions of the electric and magnetic form factors given in the dipole approximation. Then, the $Q^{2}$ integrated photon flux can be written as
\begin{eqnarray}
f(E_{\gamma}) = \int_{Q_{min}^{2}}^{Q_{max}^{2}}\frac{\mathrm d N_{\gamma}}{\mathrm dE_{\gamma}\mathrm dQ^{2}}\mathrm dQ^{2},
\end{eqnarray}
where $Q_{max}^{2} \approx 2 \sim 4$ GeV$^{2}$. Since the contribution to the above integral formula is very small for $Q_{max}^{2} > $ 2 GeV$^{2}$, in our numerical calculation, we will approximately take $Q_{max}^{2} = $ 2 GeV$^{2}$.

\begin{figure}[htb]
\vspace{-0.5cm}
\begin{center}
 \epsfig{file=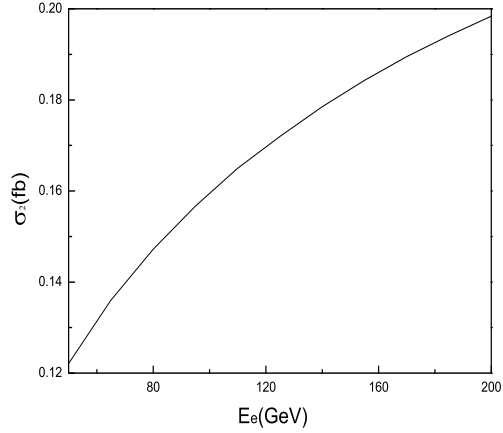, width=220pt,height=200pt}
  \vspace{-0.5cm}
  \caption{At the LHeC, the production cross section $\sigma_{2}$ of the subprocess $e \gamma \rightarrow \tau H$ as a function \hspace*{1.4cm}of the electron beam energy $E_{e}$ for $Y=\sqrt{|Y_{\tau e}|^{2}+|Y_{e \tau}|^{2}}$ = 0.014, and the detected \hspace*{1.4cm}acceptance region: 0.0015 $\leq\xi\leq$ 0.5.}
\label{ee}
\end{center}
\end{figure}

At the LHeC, the effective production cross section $\sigma_{2}(\tau H)$ for the subprocess $e \gamma \rightarrow \tau H$ can be written as
\begin{eqnarray}
\sigma_{2}(\tau H) = \int_{Max(Z, \xi_{min})}^{\xi_{max}} E \mathrm d\xi f(\xi E)\int_{(\cos\theta)_{min}}^{(\cos\theta)_{max}}\mathrm d \cos\theta \frac{\mathrm d\hat{\sigma}(\hat{s})}{\mathrm d\cos\theta},
\end{eqnarray}
where $\hat{s}$ = 4$E_{e} E_{\gamma}$ = $\xi s$ with $E_{e}$ = 50 $\sim$ 200 GeV and $E$ = 7 TeV, $Z = (m_{\tau} + m_{H})^{2}/s $. Similar as above, we also apply the cut 10$^{\circ} \leq \theta \leq 170^{\circ}$. The intact proton radiating photon can not be detected from the central detectors. However, the forward detectors can detect the particles with large pseudorapidity providing some information  on the intact proton. Based on the forward proton detectors to be installed by the CMS-TOTEM and the ATLAS collaborations, we choose the detected acceptance regions as 0.0015 $< \xi_{1} < 0.5$ and 0.0015 $< \xi_{2} < 0.15$ [24, 25, 27].

\begin{figure}[htb]
\vspace{-0.5cm}
\begin{center}
 \epsfig{file=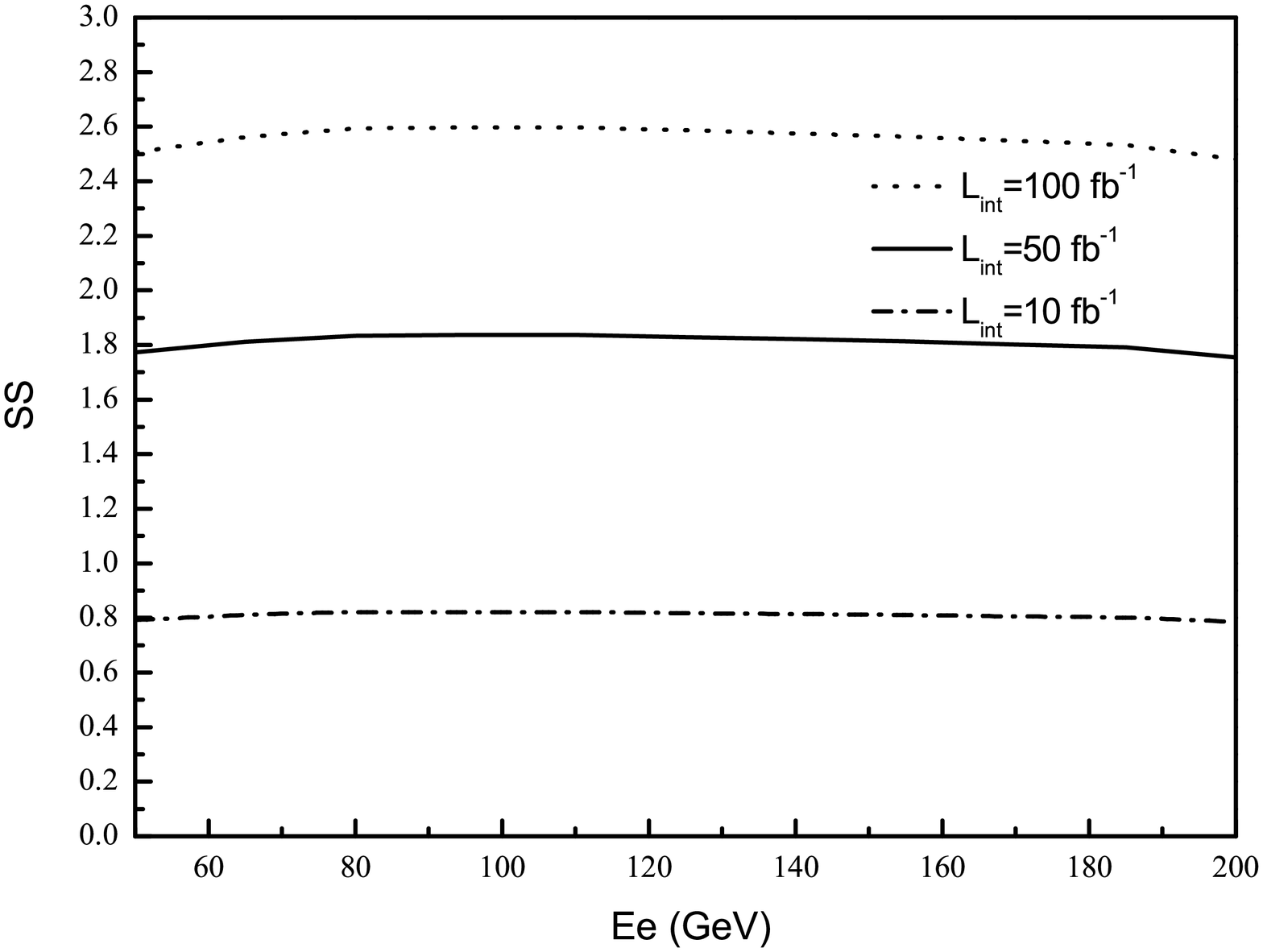, width=220pt,height=200pt}
 \vspace{-0.5cm}
  \caption{Variation of the significance $SS=S/\sqrt{S+B}$ with  the electron beam energy $E_{e}$  \hspace*{1.5cm}for different values of the integrated luminosity of the LHeC.}
 \label{ee}
\end{center}
\end{figure}

Our numerical results show that the values of the total cross section $\sigma_{2}$ of the LFV subprocess $e \gamma \rightarrow \tau H$ at the LHeC for two detected acceptance regions 0.0015 $\sim$ 0.5 and 0.0015 $\sim$ 0.15 are approximately equal to each other. The difference is smaller than $ 1\% $. Thus, in Fig.4, we only plot the total cross section $\sigma_{2}$  as a function of the electron beam energy $E_{e}$ for the values of the parameter $\xi$ in the range of 0.0015 $\sim$ 0.5. One can see that the value of the cross section $\sigma_{2}$  is in the range of 0.12 $\sim$ 0.198 fb for $E$ = 7 TeV and $E_{e}$ = 50 $\sim$ 200 GeV. Similarly with that at the ILC, the signal final state of the subprocess $e \gamma \rightarrow \tau H$ at the LHeC is $\tau b \overline{b} $. For $E$ = 7 TeV and $E_{e}$ = 50 $\sim$ 200 GeV, there only will be several $\tau b \overline{b}$ events to be generated at the LHeC experiment with the integrated luminosity $\mathcal{L}$$_{int}$ = 100 fb$^{-1}$. Though the production rate of the signal is much small,  we also use MadGraph to calculate all the signal and background events generated by $e \gamma$ collision at the LHeC, based on the basic cuts: $p^l_T>15GeV, p^b_T>20GeV, \rlap/E_T > 25GeV$, where $p_T$ denotes the transverse momentum, $\rlap/E_T$ is the missing transverse momentum from the invisible neutrino in the final state. The  significance $SS=S/\sqrt{S+B}$ are shown in Fig.5 as a function of  the electron beam energy $E_{e}$ for different values of the integrated luminosity of the LHeC. One can see from Fig.5 that, even we take the maximal value of the LFV coupling $H \tau e $, $Y = 0.014$, the value of  the  significance $SS$ is smaller than 2.5 in most of the parameter space. Thus, it is very challenge to detect  the LFV coupling $H \tau e $ via the process $e \gamma \rightarrow \tau H$ at the LHeC.

\vspace{0.5cm} \noindent{\bf 4. Conclusions}

\vspace{0.5cm}

The transition from the discovery to precision measurement of Higgs boson physics has begun. There is great potential for the SM Higgs boson as a future harbinger of new physics. The LFV Higgs couplings appear quite generally in new physics models. Low energy constraints are weak for the LFV Higgs couplings involving $\tau$ lepton, so that, with the LHC data continuing to accumulate, the LFV Higgs signals  become experimentally available.

The existence of the LFV Higgs couplings is an exciting possibility. Observation of such LFV signals in current or future experiments would provide a clear evidence of new physics beyond the SM. In this work, considering the low energy constraints on the LFV couplings $H \ell_{i} \ell_{j}$ and  the CMS results in the search for the LFV decay $H\rightarrow \mu\tau$, we study production of the SM Higgs boson via $e \gamma$ collision mediated by the LFV coupling $H \tau e $ at the ILC and LHeC experiments. Our numerical results show that the signal final state $\tau b \overline{b}$ is almost background free, which should be detected in future ILC experiments. The production cross section of the subprocess $e \gamma \rightarrow \tau H$ at the LHeC is much small, which is very challenge to be detected at the LHeC.

\section*{Acknowledgments} \hspace{5mm}This work was
supported in part by the National Natural Science Foundation of
China under Grant No. 11275088, the Natural Science Foundation of the Liaoning Scientific Committee
(No. 2014020151).

\end{document}